\documentclass[manuscript]{aastex}
\usepackage{txfonts}
\usepackage{graphicx}

\slugcomment{APRIM proceeding}

\shorttitle{Follow-up observations toward Planck cold clumps with ground-based radio telescopes} \shortauthors{Liu et al.}

\begin{document}

\title{Follow-up observations toward Planck cold clumps with ground-based radio telescopes}
\author{
Tie Liu               \altaffilmark{1,2},
,Yuefang Wu            \altaffilmark{2}
,Diego Mardones       \altaffilmark{3}
,Kee-Tae Kim         \altaffilmark{1}
,Karl M. Menten       \altaffilmark{4}
,Ken Tatematsu        \altaffilmark{5}
,Maria Cunningham     \altaffilmark{6}
,Mika Juvela          \altaffilmark{7}
,Qizhou Zhang         \altaffilmark{8}
,Paul F Goldsmith     \altaffilmark{9}
,Sheng-Yuan Liu       \altaffilmark{10  }
,Hua-Wei Zhang        \altaffilmark{2   }
,Fanyi Meng           \altaffilmark{11  }
,Di Li                \altaffilmark{12  }
,Nadia Lo             \altaffilmark{3   }
,Xin Guan             \altaffilmark{11  }
,Jinghua Yuan         \altaffilmark{12  }
,Arnaud Belloche      \altaffilmark{4   }
,Christian Henkel     \altaffilmark{4,19}
,Friedrich Wyrowski   \altaffilmark{4   }
,Guido Garay          \altaffilmark{3   }
,Isabelle Ristorcelli \altaffilmark{13  }
,Jeong-Eun Lee        \altaffilmark{14  }
,Ke Wang              \altaffilmark{15  }
,Leonardo Bronfman    \altaffilmark{3   }
,L. Viktor Toth       \altaffilmark{16  }
,Scott Schnee         \altaffilmark{17  }
,Shengli Qin          \altaffilmark{18  }
,Shaila Akhter        \altaffilmark{6   }
}

\altaffiltext{1}{Korea Astronomy and Space Science Institute 776, Daedeokdae-ro, Yuseong-gu, Daejeon, Republic of Korea 305-348; liutiepku@gmail.com}
\altaffiltext{2}{Peking University, China;yfwu.pku@gmail.com}
\altaffiltext{3}{Universidad de Chile, Chile}
\altaffiltext{4}{Max-Planck-Institut f\"{u}r Radioastronomie, Germany}
\altaffiltext{5}{ National Astronomical Observatory of Japan, Japan}
\altaffiltext{6}{The University of New South Wales, Australia}
\altaffiltext{7}{University of Helsinki, Finland}
\altaffiltext{8}{The Harvard–Smithsonian Center for Astrophysics, U.S.A.}
\altaffiltext{9}{Jet Propulsion Laboratory, U.S.A.}
\altaffiltext{10}{Academia Sinica Institute of Astronomy and Astrophysics, Taiwan}
\altaffiltext{11}{ Universit\"{a}t zu K\"{o}ln, Germany}
\altaffiltext{12}{National Astronomical Observatories, Chinese Academy of Sciences, China}
\altaffiltext{13}{IRAP-Toulouse University, France}
\altaffiltext{14}{Kyung Hee University, Korea}
\altaffiltext{15}{European Southern Observatory, Germany}
\altaffiltext{16}{Eotvos University, Hungary }
\altaffiltext{17}{The National Radio Astronomy Observatory, U.S.A.}
\altaffiltext{18}{YunNan University, China}
\altaffiltext{19}{King Abdulaziz Univ., Jeddah, Saudi Arabia}

\begin{abstract}
The physical and chemical properties of prestellar cores, especially massive ones, are still far from being well understood due to the lack of a large sample. The low dust temperature
($<$14 K) of Planck cold clumps makes them promising candidates for prestellar objects or for sources at the very initial stages of protostellar collapse. We have been conducting a series of observations toward Planck cold clumps (PCCs) with ground-based radio telescopes. In general, when compared with other star forming samples (e.g. infrared dark clouds), PCCs are more quiescent, suggesting that most of them may be in the earliest phase of star formation. However, some PCCs are associated with protostars and molecular outflows, indicating that not all PCCs are in a prestellar phase. We have identified hundreds of starless dense clumps from the mapping survey with the Purple Mountain Observatory (PMO) 13.7-m telescope. Follow-up observations suggest that these dense clumps are ideal targets to search for prestellar objects.
\end{abstract}

\keywords{stars: formation---stars: evolution---ISM: clouds---ISM: molecules}

\section{Introduction}

Stars form in dense regions within molecular clouds, called pre-stellar cores (PSCs), which provide information on the initial conditions of star formation \citep{case11}. The observations of low-mass PSCs in nearby molecular clouds have shown that PSCs are cold (T$_{k}\leq$10 K), dense (n(H$_{2}$)$>5\times10^{4}$ cm$^{-3}$), and thermally supported, with a chemistry profoundly affected by gas depletion and deuterium enrichment \citep{case11}. However, the physical and chemical properties of low-mass prestellar cores are still far from being completely understood. Do other low-mass prestellar cores share similar properties with the well-studied ones? How do different environments and external conditions affect the properties and structures of PSCs? Additionally, the number of well-studied low-mass PSCs is still very limited. Thus, the physical and chemical properties of low-mass prestellar cores need to be investigated systematically with an appropriately large sample.

The initial conditions of high mass star formation are even less well determined due to a lack of a proper sample of relatively nearby massive PSCs. The search for massive PSCs is very challenging because of their large distances, short dynamical time scales and feedback (e.g. outflows) from neighbouring active sites of high-mass star formation. Candidates for massive PSCs have been found in recent observations, e.g. G028 C1-S \citep{tan13}. When compared to low-mass prestellar cores, which are thermally supported, these candidates are characterized by a supersonic non-thermal velocity dispersion and their internal pressure is provided by a combination of turbulence and magnetic field \citep{tan13}. However, whether these candidates are really massive is still under debate due to large uncertainties in their mass estimates. Additionally, these candidates are usually at large distances (5 kpc for G028 C1-S) and may further fragment into sub-cores, a possibility that can be checked only with higher angular resolution observations. Do massive PSCs really exist? What is the difference between low-mass and high-mass prestellar cores? To answer these questions, we need to identify a large sample of massive PSCs candidates.

The Planck Early Cold Cores Catalog (ECC) provides an unbiased list of 915 reliable Galactic cold
clumps with low dust temperature ($<$14 K), which are likely pre-stellar objects or at the earliest stages of protostellar collapse \citep{planck11}. Thus, these Planck cold clumps (PCCs) are ideal targets to study the initial conditions of star formation.

\section{Observations}

\begin{table*}[t!]\small
\caption{Summary of the observations}
\centering
\begin{tabular}{cccc}
\cline{1-4}
Telescopes        & Tracers & Aims & Status \\
\cline{1-4}
\cline{1-4}
\textbf{PMO 13.7-m}       &  J=1-0 of $^{12}$CO,$^{13}$CO, C$^{18}$O,  & dense clumps  &  674 PCCs surveyed     \\
                 &  HCN and HCO$^{+}$ &              &   \\
\textbf{CSO 10-m}         &  J=2-1 of $^{12}$CO,$^{13}$CO, C$^{18}$O  &  CO depletion \& outflows & 20 PCCs mapped      \\
\textbf{APEX 12-m}        & J=2-1 of $^{12}$CO,$^{13}$CO, C$^{18}$O   & CO depletion \& outflows   &  proposal accepted \\
\textbf{NANTEN2 4-m}        &  $^{12}$CO (4-3) \& (7-6)                 &  shock evidence             &  proposal accepted \\
\textbf{IRAM 30-m}        &  J=2-1 of $^{12}$CO,$^{13}$CO \& C$^{18}$O  &   starless cores;  & 24 PCCs mapped     \\
                 & J=1-0 of HCN, HCO$^{+}$ \& N$_{2}$H$^{+}$  & chemistry &\\
\textbf{Mopra 22-m}            & J=1-0 of HCN, HCO$^{+}$ \& N$_{2}$H$^{+}$  & chemistry & 30 PCCs mapped     \\
                 & SiO (2-1), HC$_{3}$N (10-9)                &               & \\
\textbf{Effelsberg  100-m}  & NH$_{3}$ (1,1) \& (2,2);   & kinetic temperature \& chemistry  & proposal accepted      \\
                           & HC$_{7}$N (21-20) & &\\
\textbf{The SMA}   & continuum and molecular lines  & fragmentation \& kinematics  &  proposal accepted \\
                   & at 230 GHz band                & of G207.3-19.8 & \\
\cline{1-4}
\cline{1-4}
\end{tabular}
\end{table*}

To study the physical and chemical properties of PCCs and to build a large sample for candidates of prestellar objects, we have been conducting
a series of observations toward PCCs with ground-based radio telescopes, which are summarized in Table 1. In the next section, we will present some preliminary results of this
large on-going project.

\section{Results}
\subsection{Preliminary results of observations with the Purple Mountain Observatory (PMO) 13.7-m telescope}
A survey toward 674 Planck cold clumps of the Early Cold Core
Catalogue (ECC) in the J=1-0 transitions of $^{12}$CO, $^{13}$CO and
C$^{18}$O has been carried out using the PMO 13.7 m telescope since 2011 \citep{wu12,liu12,liu13,meng14}.
All clumps but one were detected in the $^{12}$CO and $^{13}$CO, and $68\%$ of
them in the C$^{18}$O line. A high consistency of the three line peak velocities was revealed
toward most of PCCs, indicating that there are no bulk motions inside them. The kinematic distances of these PCCs range from 0.1 to 21.6 kpc. 82\% of them
are located within 2 kpc and 51\% of them between 0.5 and 1.5 kpc. Excitation temperatures of $^{12}$CO (1-0) range from 4 to 27 K. Column densities
N$_{H_{2}}$ range from 10$^{20}$ to 4.5$\times10^{22}$ cm$^{-2}$ with
an average value of (4.4$\pm$0.2)$\times10^{21}$ cm$^{-2}$. The mean line widths of the J=1-0 transitions of $^{12}$CO, $^{13}$CO and
C$^{18}$O are 2.03$\pm$0.05, 1.27$\pm$0.03, and 0.76$\pm$0.03 km~s$^{-1}$ respectively. We find that most of the clumps have
non-thermal velocity dispersions ($\sigma_{NT}$) larger than thermal velocity dispersions ($\sigma_{Therm}$), indicating that these PCCs are most likely turbulence dominated.
When compared to infrared dark clouds (IRDCs), PCCs have much smaller linewidths and are more quiescent, suggesting that PCCs may be in an earlier phase in cloud evolution than IRDCs \citep{wu12}.

By combining the dust emission data from the Planck survey with the PMO 13.7 m molecular line data, we also investigate the CO abundance, CO depletion and CO-to-H$_{2}$ conversion factor \citep{liu13}. The median and mean values of the CO abundance are 0.89$\times10^{-4}$ and 1.28$\times10^{-4}$, respectively. The mean and median CO depletion factors are 1.7 and 0.9, respectively, indicating that most of PCCs may be in an earlier evolutionary phase than prestellar cores which have a much larger CO depletion factor. The median value of $X_{CO-to-H_{2}}$ for the whole sample is $2.8\times10^{20}$ cm$^{-2}$K$^{-1}$km$^{-1}$~s. The CO abundance, CO depletion factor and CO-to-H$_{2}$ conversion factor are strongly (anti-)correlated to other physical parameters (e.g. dust temperature, dust emissivity spectral index, column density, volume density and luminosity-to-mass ratio) \citep{liu13}.

We have also mapped $\sim$500 PCCs in the same transitions with the PMO 13.7-m telescope and the survey is still on-going. From the mapping survey, more than 700 dense clumps have been identified and follow-up observations toward these dense clumps have been proposed as shown in Table 1.

\subsection{Not all PCCs are starless}
PCCs are believed to be likely pre-stellar objects or targets at the earliest stages of protostellar collapse \citep{planck11}. However, from the CSO observations, we realized that some PCCs are in a more evolved phase than prestellar cores. In Figure 1, we show two PCCs (G177.1-01.2 \& G192.3-11.8) which are associated with molecular outflows. The WISE images indicate that G177.1-01.2 is associated with two protostars. The C$^{18}$O (2-1) emission peak is offset from the protostars, indicating that the C$^{18}$O core may be starless, which needs to be confirmed by further higher spatial resolution observations. A bipolar outflow is revealed in $^{12}$CO (2-1) emission. The outflow seems to be driven by the southern protostar. G192.3-11.8 is associated with a faint brown dwarf \citep{Gri12}, which is marked by a yellow cross in the lower-middle panel of Figure 1. The C$^{18}$O (2-1) emission reveals a small nebula around the brown dwarf. From the $^{12}$CO (2-1) emission, we find that the brown dwarf drives a slow ($\sim$2 km~s$^{-1}$) and compact bipolar outflow.

Although some PCCs seem to be protostellar objects, we have already identified hundreds of starless dense clumps. In the right panels of Figure 1, we present two starless PCCs. They are gravitationally bound dense clumps seen in the $^{13}$CO (1-0) emission. There are no associated point sources in the WISE images nor in the IRAS and MSX catalogues, indicating that they are most likely starless. Such starless dense clumps are ideal targets to search for prestellar objects.

\subsection{Search for prestellar objects}

We have been carrying out follow-up observations toward those starless dense clumps to search for prestellar core candidates. From IRAM 30-m observations, we have identified several tens of candidates of prestellar objects. In Figure 2, we present a sample of the IRAM 30-m observations. G207.3-19.8 is located north-west of the Orion nebular cluster (``ONC"). From HCO$^{+}$ (1-0) emission, we have identified several gravitational bound starless cores, which are good candidates for prestellar cores. As shown in the left panel of Figure 2, the starless cores (marked with yellow crosses) are not associated with any red infrared point sources in the WISE three-color composite image. In contrast, the protostellar core (HH 58) is associated with a red point source. As shown in the second panel, we also found a highly fragmented massive filament containing a chain of starless cores. The total mass of the filament is $\sim$300 M$_{\odot}$. From the Moment 1 map of HCO$^{+}$ (1-0) in Figure 2, we find a large velocity gradient along the filament. We estimate a velocity gradient of $\sim$3 km~s$^{-1}$~pc$^{-1}$, which is comparable to or larger than that observed in IRDCs, indicating possible flows along this filament. From the Moment 2 map of HCO$^{+}$ (1-0), we find that the filament is characterized by a supersonic non-thermal velocity dispersion, which is different from nearby low-mass and thermally supported prestellar cores. Especially, from the Moment 2 map of HCO$^{+}$ (1-0), one can see that the central part of the filament has smaller velocity dispersion ($\sim$0.6 km/s) than that of the edge or surroundings, indicating that turbulence may dissipate in the inner part of the filament.

From the IRAM 30-m observations, we have also identified six massive and starless clumps, which have masses and source-averaged column densities larger than 80 M$_{\odot}$ and $\sim$10$^{22}$ cm$^{-2}$, respectively. These six objects are ideal targets to search for candidates of pre-stellar clusters and massive prestellar cores. Our follow-up observations (e.g. Effelsberg 100-m and the SMA) will confirm whether they are real massive prestellar objects or not.

From previous observations, we have learned that PCCs are excellent targets to search not only for low-mass but also for high-mass prestellar objects. From follow-up observations, we aim to build a large sample of prestellar objects and systematically study their physical and chemical properties.

\begin{figure*}[t]
\centering
\includegraphics[angle=-90,width=140mm]{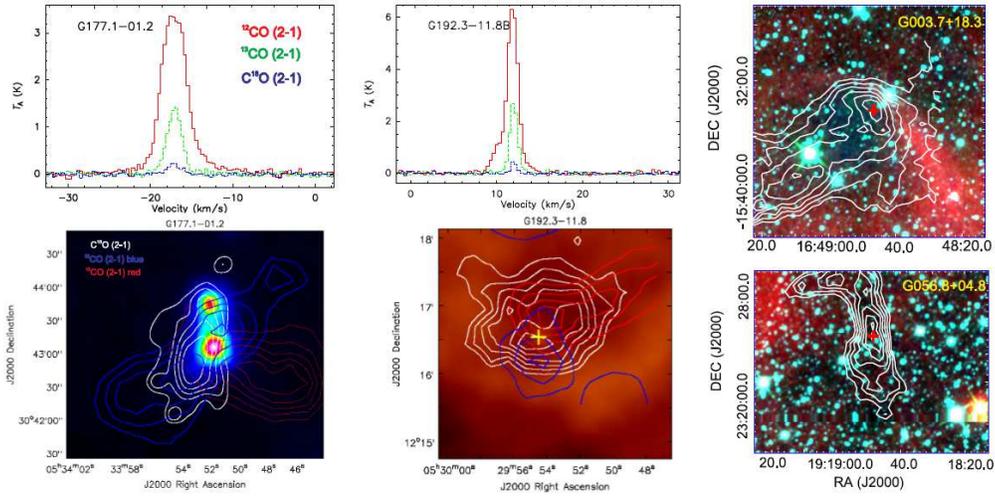}
\caption{The red, green and blue spectra in the upper-left and upper-middle panels show the clump-averaged $^{12}$CO (2-1), $^{13}$CO (2-1) and C$^{18}$O (2-1) lines, respectively. In the lower-left and lower-middle panels, the background images are WISE band w4 (22 $\mu m$) emission, and the white, blue, and red contours  represent the integrated intensity maps of C$^{18}$O (2-1), high-velocity blueshifted and redshifted emission of $^{12}$CO (2-1), respectively. The white contours in the right panels represent column density maps of $^{13}$CO (1-0) and the background image is a WISE three-color composite image (band w1 (3.4 $\mu m$) in blue, band w2 (4.6 $\mu m$) in green and band w4 (22 $\mu m$) in red). \label{fig:pkasfig1}}
\vspace{5mm} 
\end{figure*}

\begin{figure*}[t]
\centering
\includegraphics[angle=-90,width=170mm]{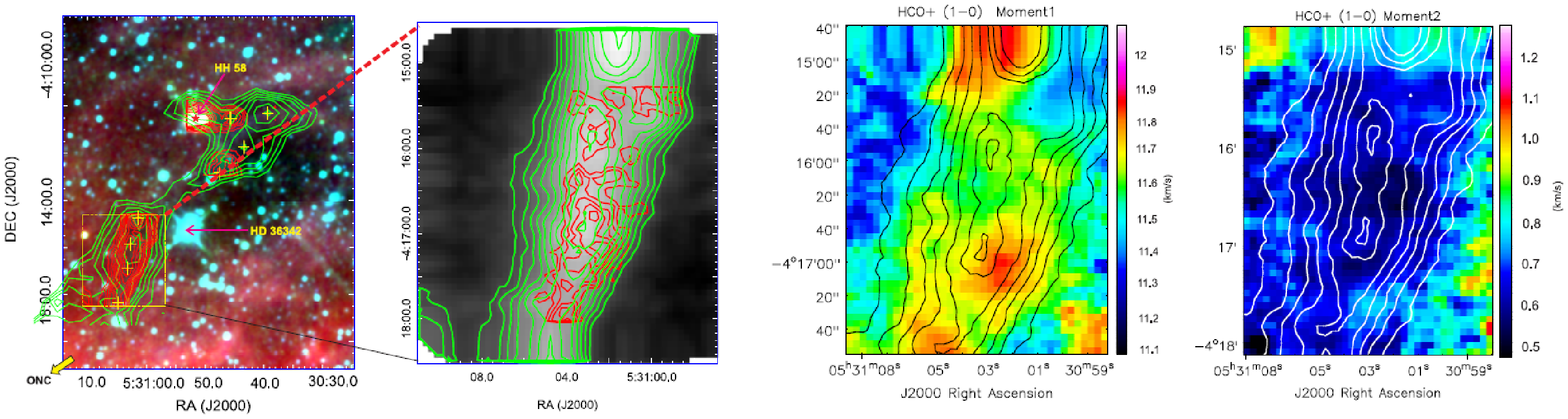}
\caption{Planck cold clump G207.3-19.8. From left to right, first panel: The background image is a WISE 3 color composite image (band w1 (3.4 $\mu m$) in blue, band w2 (4.6 $\mu m$) in green and band w4 (22 $\mu m$) in red). The two bright stars ``HH 58" and ``HD 36342" in this field are marked with arrows. The famous Orion Nebula Cluster (``ONC") is south-east of this field. The green and red contours represent integrated intensity maps of $^{13}$CO (1-0) and HCO$^{+}$ (1-0), respectively. Second panel: The green contours and background grey image represent integrated intensity of HCO$^{+}$ (1-0), while the red contours represent integrated intensity maps of N$_{2}$H$^{+}$ (1-0). Third panel:  The contours and background color image represent integrated intensity and intensity weighted velocity (Moment 1) of HCO$^{+}$ (1-0), respectively. Fourth panel: The contours and background color image represent integrated intensity and intensity weighted velocity dispersion (Moment 2) of HCO$^{+}$ (1-0), respectively.\label{fig:pkasfig2}}
\vspace{5mm} 
\end{figure*}


\acknowledgments

We are grateful to the staff at the Qinghai Station of PMO, the CSO and the IRAM 30-m for their
assistance during the observations. Yuefang Wu  and Tie Liu thank for the
support of the China Ministry of Science and Technology under State Key Development
Program for Basic Research (2012CB821800), the grants of NSFC
number 11373009 and 11433008.


{}


\end{document}